\documentstyle[aps,epsfig,prl,twocolumn]{revtex}
\draft
\newcommand{\be}{\mbox{e}}
\newcommand{\bt}{\mbox{\bf t}}
  
\newcommand{\bolT}{\mbox{\bf T}}
\begin{document}
\date{19 June 2000}
\twocolumn[\hsize\textwidth\columnwidth\hsize\csname
@twocolumnfalse\endcsname
\title{Low-Lying Magnetic Excitation of the Shastry-Sutherland
Model}
\author{K.~Totsuka$\mbox{}^{\ast}$}
\address{Department of Physics, Kyushu University, 
6-10-1 Hakozaki, Fukuoka, 812-8581, Japan} 
\author{S.~Miyahara, and K.~Ueda}
\address{Institute of Solid State Physics, University of Tokyo, 
5-1-5 Kashiwanoha, Kashiwa, Chiba, 277-8581, Japan} 
\maketitle
\begin{abstract}
By using perturbation calculation and numerical diagonalization,
low-energy spin dynamics of the Shastry-Sutherland model
is investigated paying particular attention to the two-particle
coherent motion.  
In addition to spin-singlet- and triplet bound states,  
we find novel branches of coherent motion
of a bound quintet pair, which are usually unstable because
of repulsion.  
Unusual dispersion observed in neutron-scattering 
measurements are explained by the present theory.   
The importance of the effects of phonon is also pointed out. 
\end{abstract}
\begin{center}
\end{center}
]
\narrowtext
In recent years, low-dimensional spin systems with a spin gap
have been a subject of extensive research.
Among them, a two-dimensional antiferromagnet
$\mbox{SrCu}_{2}(\mbox{BO}_{3})_{2}$ is outstanding in its
unique features.   These include, (i) spin-gapped behavior
\cite{Kageyama-Y-S-M-O-K-K-S-G-U-99},
(ii) magnetization plateaus \cite{Onizuka-K-N-K-U-G-00},
and (iii) unusual low-energy dynamics
\cite{Kageyama-N-A-O-Y-N-K-K-U-00}.
In Ref.\cite{Miyahara-U-99}, it was pointed out that
$\mbox{SrCu}_{2}(\mbox{BO}_{3})_{2}$ may be modeled by
the $S=1/2$ Heisenberg model on the Shastry-Sutherland (SS) lattice
\cite{Shastry-S-81} (Shastry-Sutherland model, hereafter).
This sparked experimental- and theoretical researches on
interesting features of the Shastry-Sutherland model
\cite{Weihong-H-O-99,Nojiri-K-O-U-M-99,%
Momoi-T-99,Miyahara-U-00,M-Hartmann-S-K-U-99,Koga-K-00}.

Strong geometrical frustration of the SS model
allows a simple dimer-product to be the exact ground state
\cite{Shastry-S-81}.      In the zeroth-order approximation,
a triplet excitation above the dimer-singlet ground state
is created by promoting one of the dimer singlets to triplet.
In this letter, we consider such particle-like excitations and
show that interesting two-particle motion (bound states)
is possible because of unusual dynamical properties.

The unit cell of the SS lattice contains two mutually
orthogonal dimer bonds (we call them A and B; see Fig.1) and
the Hamiltonian can be written as a sum of local Hamiltonians
acting only on either A- or B dimers and those acting on both
A and B dimers:
\begin{equation}
{\cal H} = \sum_{\alpha = {\rm A,B}}{\cal H}_{\alpha}
+ \sum_{(\alpha,\beta)=\langle {\rm A,B}\rangle }
{\cal H}_{\alpha,\beta} \; .
\end{equation}
In terms of the hardcore triplet bosons $\bt$ on
dimer bonds \cite{Sachdev-B-90},
the local Hamiltonians are given by
${\cal H}_{\alpha} = J[-\frac{3}{4} + \bt_{\alpha}^{\dagger}{\cdot}
\bt_{\alpha}]$ and
${\cal H}_{\alpha,\beta} =
\frac{J'}{2}[ \mbox{sig}(\alpha,\beta) \{
i \bt^{\dagger}_{\alpha}{\cdot}(\bt_{\beta}\times\bt_{\alpha})
+ (\mbox{h.c.})\} + \bolT_{\alpha}{\cdot}\bolT_{\beta} ] $,
where $\bolT_{\alpha,\beta}$ denote the $S=1$ operators and
the sign factor $\mbox{sig}(\alpha,\beta)$ equals 1 when
the arrow on a horizontal bond ($\beta$) is emanating from
the vertical one ($\alpha$) and $-1$ otherwise (see Fig.1).

As is easily seen, a unique geometry of the SS lattice
allows neither (bare) one-particle hopping
($\bt^{\dagger}(x){\cdot}\bt(y)$) nor pair
creation/annihilation of triplets;
non-trivial one-particle (triplet) hopping
is generated only perturbatively \cite{Miyahara-U-99}
(it occurs at $(J'/J)^{6}$ and higher).

\noindent%
\underline{{\em Correlated hopping}}\\
Although one-particle hopping is strongly suppressed,
the situation is dramatically different for
two-particle cases.
The 3-point vertices (e.g. $\bt^{\dagger}{\cdot}(\bt\times\bt)$)
contained in the Hamiltonian make
non-trivial two-particle hopping like Fig.~2 possible
already at $(J^{\prime}/J)^{2}$.
Note that only one of the
two particles hops and that the other is at rest  
merely to assist the hopping.
Therefore, we call such processes {\em correlated hoppings}; 
two triplets close to each other can use this 
{\em new} channel of two-particle motion to form various 
bound states.  
Although the relevance of correlated hoppings in  
the SS model was pointed out in \cite{Momoi-T-99}
in the context of magnetization process,  
the effects of it would be most highlighted in 
the low-energy dynamics.

Usually, correlated motions are only higher-order
corrections to the dominant one-particle processes.
However, if the one-particle processes are strongly
suppressed for some reasons, correlated ones would
play an important role in the low-energy dynamics.
The Shastry-Sutherland model is a candidate for such systems.
Below we demonstrate that the correlated hoppings can
explain the unusual dispersion observed in
inelastic neutron-scattering (INS) experiments
\cite{Kageyama-N-A-O-Y-N-K-K-U-00}.

\noindent%
\underline{{\em Two-triplet motion}}\\
Collecting all the two-particle processes
up to $(J^{\prime}/J)^{3}$,
we found that the two-particle coherent motion is closed
within the four states (Fig.~2) which are decoupled from other
states where two triplets are far apart.   
Because of this special property, if we take the following 
four relative configurations (Fig.~2) 
\begin{equation}
| \Psi(p_{x},p_{y}) \rangle =
(|a(\mbox{\bf p})\rangle, |b(\mbox{\bf p})\rangle,
|c(\mbox{\bf p})\rangle, |d(\mbox{\bf p})\rangle ) 
\end{equation}
as the basis, computation of the two-triplet 
spectra reduces to diagonalization of a four-by-four (up to 
3rd order) hopping matrix.   
Although the form of it is the same as 
that given in Ref.~\cite{Momoi-T-unpub-00}, 
the elements now  
are functions of spin-1 operators $\bolT_{1}$ and $\bolT_{2}$
of the two triplets; for example, the interactions
$V_{\rm NN}$ and $V_{\rm NNN}$
between nearest-neighbor- and next-nearest-neighbor pair
are given by
\begin{eqnarray}
& & V_{\rm NN}=\left( \frac{1}{2}J^{\prime}
-\frac{(J^{\prime})^{2}}{4J} -\frac{(J^{\prime})^{3}}{2J^{2}} \right)
\bolT_{1}{\cdot}\bolT_{2} \nonumber \\
& & \;\;\;\; \;\;\;\;- \left(
\frac{(J^{\prime})^{2}}{4J} + \frac{(J^{\prime})^{3}}{8J^{2}}
\right) (\bolT_{1}{\cdot}\bolT_{2})^{2}
+ \left( \frac{(J^{\prime})^{2}}{J} +
\frac{(J^{\prime})^{3}}{2J^{2}} \right)
\nonumber \\
& & V_{\rm NNN} = \frac{(J^{\prime})^{3}}{4J^{2}}
\bolT_{1}{\cdot}\bolT_{2} 
\; .
\end{eqnarray}
The meaning of the hopping amplitudes $J_{\rm NN}$ and
$J_{\rm 3rd}$ can be read off from Fig.~2.
Note that
$\mbox{\bf p}$ is defined with
respect to the chemical unit cell
differently
from that used in Refs.\cite{Kageyama-N-A-O-Y-N-K-K-U-00,%
Weihong-H-O-99}.
For completeness, we add another two-particle interaction
between a 3rd-neighbor pair ($|\cdots|$):
\begin{equation}
V_{\rm 3rd}=
\left( \frac{(J^{\prime})^{2}}{2J}
+ \frac{3(J^{\prime})^{3}}{4J^{2}}\right)
\bolT_{1}{\cdot}\bolT_{2}  \; ,
\end{equation}
which creates immobile (at this order) bound pairs with energy
$2V_{0}(J,J^{\prime}) +V_{\rm 3rd}$ for $S_{\rm tot}=0, 1$.
In the perturbative regime, a pair with total spin
$S_{\rm tot}=0$ or 1 feels
attraction whereas one with $S_{\rm tot}=2$
repels each other, which is the origin of magnetization
plateaus of the SS model \cite{Miyahara-U-00,Momoi-T-99}.
The occurrence of attraction for a singlet- or triplet pair is
not restricted to the SS system and
indeed is responsible for the bound states in
the 2-leg ladder \cite{Sushkov-K-98} and the $S=1/2$ dimerized
chain \cite{Barnes-R-T-99}.
Interchange of the sub(dimer)lattices A$\rightarrow$B 
($(p_{x},p_{y})\mapsto (p_{y},-p_{x})$) gives another 
hopping matrix corresponding to different bound states.  
Diagonalizing them, we obtain 8 branches
(4 for each dimer sublattice) for a given value of total spin. 
Of course, on top of them,
there are dispersionless bound states corresponding to
$V_{\rm 3rd}$ and infinitely degenerate (up to 5th order)
levels corresponding to two isolated triplets at $\omega =
2V_{0}(J,J^{\prime})$.  At higher orders of perturbation,
the infinite degeneracy will be lifted and this level becomes
a narrow continuum.
The dispersion curves obtained this way are shown in Fig.~3-5 
for $J^{\prime}/J=0.5$.   
Note that the entire spectrum is invariant under
$\mbox{D}_{\rm 2d}$-operations
$(p_{x},p_{y}) \mapsto (p_{y},p_{x})$, $\quad
(p_{x},p_{y}) \mapsto (-p_{x}, p_{y})$ and their products.

As is expected, there appear stable branches of bound states
(bold solid lines and dotted lines)
below the two-particle threshold (thin broken lines) for
$S_{\rm tot}=0$ and 1.   Because of a strong attraction
between two triplets on adjacent dimer bonds,
the energy decrease from the two-particle threshold is largest for
the singlet bound states (Fig.~3).  
As a direct consequence of the correlated hopping starting 
from $(J^{\prime}/J)^{2}$, the dispersion of the bound states 
is relatively large compared with that of the lowest 
single-triplet excitation, which is less than 
$0.01J$ for $J'/J=0.5$.    
To supplement the perturbation theory (PT)
we also performed exact diagonalization (ED) of the
original Hamiltonian for finite clusters  
with 16 and 24 sites 
under periodic boundary conditions.  
For small $J'/J$, e.g. 0.2, we have confirmed that the results 
of ED for $S_{\rm tot} = 0$ and 1 are consistent with those 
of PT.    

We investigated excited states for larger $J'/J$ as well.
While the bandwidth becomes larger for most branches,
an almost flat band was found even for relatively large $J'/J$.  
For small $J'/J$, the energy obtained
by ED is close to that of the 3rd-neighbor bound pairs 
($2V_{0} + V_{\text{3rd}}$) calculated by PT.   
Therefore we may identify this flat band with  
the aforementioned 3rd-neighbor bound pairs.  

What is more interesting is that although a quintet pair feels
repulsion several branches lie slightly below the two-particle
threshold.    
This binding beginning at $(J'/J)^{3}$  
is of purely dynamical origin and is peculiar to 
the orthogonal dimer systems.   
Its physical implications to magnetization process 
are discussed in detail in Ref.~\cite{Momoi-T-unpub-00}.    
For $S_{\rm tot} = 2$ it is difficult to compare
the results of ED with those of PT,
because there are so many levels around the two-particle threshold.
However, the minimum energy lies below the two-particle threshold
as is suggested from PT.   
We examined numerical data for various $J'/J$ and found that  
this seems to be the case at least up to $J'/J \approx 0.55$,
beyond which no decisive conclusion was drawn due to
finite-size effects.

Quite recently, Nojiri {\em et al.} found  in the ESR (electron 
spin resonance) spectra 
a quintet branch at about 1400GHz,
which is slightly smaller than twice the single triplet gap
722GHz \cite{Nojiri-unpub-00}.
We believe that our findings are of direct relevance
to this observation.   

\noindent%
\underline{{\em Selection Rules}}\\
The unique structure of the SS lattice allows  
only a few branches to be observed  
in INS \cite{Kageyama-N-A-O-Y-N-K-K-U-00}.    
The Fourier transform $S^{j}(\mbox{\bf p})$
$(j=x,y,z)$ of local spin operators create single-dimer-triplet 
states 
\begin{equation}
S^{j}(\mbox{\bf p}) |\mbox{G.S.}\rangle
 =  f_{+}(\mbox{\bf p}) |j; \mbox{A}( \mbox{\bf p}) \rangle +
f_{-}(\mbox{\bf p})
\be^{-i \frac{l}{2}(p_{x}+p_{y})}
|j; \mbox{B}( \mbox{\bf p}) \rangle
\end{equation}
over the dimer ground state, where 
$|j; \mbox{A/B}( \mbox{\bf p}) \rangle
\equiv t^{\dagger}_{\text{A/B},j}(\mbox{\bf p})
|\mbox{G.S.}\rangle$.   
Since one of the structure factors
$f_{\pm}(\mbox{\bf p})
=\mp i \sin ( l_{\rm d} (p_{x}\pm p_{y})/2\sqrt{2} )$ vanishes on
$p_{x}=\mp p_{y}$,
only an A (B) triplet is excited along the line $\Sigma$:
$p_{x}=p_{y}$ ($p_{x}=-p_{y}$).  
Although the state $S^{j}(\mbox{\bf p})|\mbox{G.S.}\rangle$
contains only a single triplet,
the first-order perturbation (3-point vertices)
broadens the wave function and
$S^{j}(\mbox{\bf p})|\mbox{G.S.}\rangle$
can have a finite overlap with the two-triplet states.   
Since $S^{j}(p_{x},\pm p_{y})|\mbox{G.S.}\rangle$ is even under 
the reflection $\sigma_{\rm d}(1,\pm )$ 
(i.e. belongs to $\Sigma_{1}$-representation), 
any states which are connected to  
$S^{j}(p_{x},\pm p_{y})|\mbox{G.S.}\rangle$  
by perturbation should also be even.   
From this, it follows that any bound-state wave functions 
which contain ``$b$" and ``$c$" in a {\em antisymmetric} manner
(i.e. $\Sigma_{2}$) are orthogonal to
$S^{j}(\mbox{\bf p})|\mbox{G.S.}\rangle$;
that is, INS experiments performed
along the [110] ([1$\bar{1}$0])) direction
(as in Ref.\cite{Kageyama-N-A-O-Y-N-K-K-U-00})
observe only bound states shown by solid black lines  
in Fig.~4.     
Qualitative features agree with what was observed in experiments 
\cite{Kageyama-N-A-O-Y-N-K-K-U-00}. 

To compare our results with experiments quantitatively,  
we carried out ED since the value $J'/J=0.635$   
recently estimated \cite{Miyahara-U-00-2}  
for $\mbox{SrCu}_{2}(\mbox{BO}_{3})_{2}$ is too large 
for the results of PT to be used.    
The results for $J=85$K are shown in Fig.~6.   
By close inspection of the wave function obtained both by PT and 
ED, we can identify the INS-active branches 
shown by the lower black lines in Fig.~4 
in the numerical spectrum for $J'/J=0.635$.    
The energies thus obtained are plotted by open circles 
in Fig.~6 giving good agreement with the experimental 
value (filled circles) \cite{Kageyama-N-A-O-Y-N-K-K-U-00}.    
In particular, the energy of the lowest INS-active 
two-triplet is given by 4.99meV at {\bf p}=0.   
This agrees with 5.0meV observed  
in experiments \cite{Kageyama-N-A-O-Y-N-K-K-U-00} 
and supports the validity of the value $J'/J=0.635$ for 
$\mbox{SrCu}_{2}(\mbox{BO}_{3})_{2}$.    

This selection rule does not exclude the possibility of
observing the remaining branches by other methods.
For example, optical methods probe excitations
at $\mbox{\bf p}=0$.  
The representations (E, $\mbox{B}_{2}$) and  
($\mbox{A}_{1}$, $\mbox{A}_{2}$, 
$\mbox{B}_{1}$, $\mbox{B}_{2}$, E) are active in 
far infrared spectroscopy (FIR) \cite{Room-N-L-K-O-U-99} and 
Raman scattering \cite{Lemmens-G-F-G-K-K-O-U-00}, 
respectively.   
Analysis of the wave functions shows that, for example,
the 1-triplet, the lowest 2-triplet (singlet pair),
and the second-lowest 2-triplet (3rd-neighbor singlet pair)
in Fig.~3, 4 belong to  
E, $\mbox{A}_{2}\oplus\mbox{B}_{1}$,  
and $\mbox{A}_{1}\oplus\mbox{B}_{2}$, respectively.
This is qualitatively consistent with the results of the experiments.  
Moreover, analytical- and numerical results suggest that  
the lowest bound state in the triplet sector is dominated by 
the 3rd-neighbor pair (E-rep), which is not excited 
by $S^{j}(\mbox{\bf p})$; using the above parameters, 
the energy of it (at {\bf p}=0) is given by 36.5$\mbox{cm}^{-1}$, 
which does not contradict the FIR- and ESR results.  
Detailed analyses will be published elsewhere.

\noindent%
\underline{{\em Phonons}}\\
Up to now, we have treated dynamics of a purely magnetic system.
In real materials, coupling between phonons and 
spin degrees of freedom would be important.  
To see how inclusion of virtual phonons yields effective
interactions,
we treat only the simplest case of the Einstein phonon
where each dimer ($J$) bond rotates independently around
its equilibrium position with a frequency $\omega_{0}$.
In contrast to the purely magnetic case, pair
creation/annihilation and one-particle hopping which accompany
emission of a phonon are allowed.

Here we only consider the most important case of
a nearest-neighbor pair.
A little algebra shows that the second-order processes generate
an attractive interaction $V^{\rm (ph)}_{\rm NN} \propto
-(2J+\omega_{0})/(J+\omega_{0})$ even for a quintet pair;
a singlet pair feels attraction while
a triplet repulsion.
Therefore, we may expect that the lowering in energy for quintet
bound states is enhanced by phonons.

In $\mbox{SrCu}_{2}(\mbox{BO}_{3})_{2}$ 
the interlayer coupling $J_{\perp}$ is not negligible either.  
However, $J_{\perp}$ changes nothing as far as  
{\em intra}-layer excitations are considered \cite{Miyahara-U-00-2}.  
It might make a new (INS-inactive) 
interlayer bound states slightly below 
the two-particle threshold by a very weak 
interaction $J_{\perp}(1-1/4(J'/J)^{2})\mbox{\bf T}_{1}{\cdot}
\mbox{\bf T}_{2}$.  These bound states 
have a small bandwidth of the order ($(J'/J)^{6}$).  

In conclusion, we demonstrated that unique dynamical property of
the SS model allows an interesting pairing of triplets.
Bound two triplets can move on a lattice much easier than 
a single isolated triplet can and this fact explains the
recent INS experiments.   Note that the difference in dispersion 
is not quantitative and originates from that in hopping mechanisms.    
Moreover, it enables the two triplets to form quintet  
`bound' states despite the repulsion between them.    
To complement the perturbative
consideration, we also carried out exact diagonalization
and obtained for $J'/J=0.635$
the results consistent with those of experiments.  

After this work was completed, we became aware of 
a preprint by Knetter {\em et al} (cond-mat 0005322).  
They investigated gaps at {\bf p}=0 for singlet- and triplet bound states 
using a similar but different method.  

\begin{acknowledgments}
The authors are grateful to H.~Kageyama, N.~Kawakami,
P.~Lemmens,
and H.~Nojiri for discussions and showing their unpublished
results.  The author (K.T.) thanks A.~Kolezhuk, H.~Kuroe and
T.~Momoi for stimulating discussions.
S.M. was financially supported by JSPS Research
Fellowships for Young Scientists.
\end{acknowledgments}

\begin{center}
{\bf CAPTIONS}
\end{center}
\noindent%
{\bf Fig.~1} \\
Configurations of two orthogonal dimers 
in a unit cell.  
If the direction of the vertical ($\alpha$) bond is fixed
(say, pointing downward), two configurations (left and right) 
are allowed for the horizontal ($\beta$) bond. \\
{\bf Fig.~2}\\
``Hopping'' processes of two triplets.
A(B) dimers are shown by bold black (gray) lines.
A filled circle denotes a representative point of
a unit cell {\bf r}. Symbols for the matrix elements are the same 
as those in Ref.~\cite{Momoi-T-unpub-00}. \\
{\bf Fig.~3} \\
Singlet ($S_{\text{tot}}=0$)
dispersion in the [110]- and [100]-direction for 
$J'/J=0.5$ obtained by the perturbation.  
Dashed- and dotted lines denote B-branches and 3rd-neighbor  
bound states, respectively.   Two-particle threshold lies at 
$1.356J$ (thin broken line).  \\
{\bf Fig.~4} \\
Same for triplet ($S_{\text{tot}}=1$) sector.
Only branches shown by solid black lines are observable
in neutron-scattering experiments.  \\ 
{\bf Fig.~5} \\
Same for quintet ($S_{\text{tot}}=2$) sector.
Note that several branches lie below 2-particle threshold
(thin broken line).  Because of higher-order processes, 
states above the 2-particle threshold 
(thin dashed line) are unstable.  \\
{\bf Fig.~6} \\
The branches with $S_{\text{tot}}=1$.
The results of the ED 
for $J'/J = 0.635$ and $J = 85$ K
are shown by open symbols.
Among them the branches observable 
in neutron-scattering experiments 
are shown by the open circles (lower: 1-triplet, 
upper: 2-triplet).
For comparison, the experimental results are shown
by the closed circles.  
Small splittings at {\bf p}=0 are artifacts coming from the shape 
of a cluster (which is $C_{2}$-invariant) used in ED.  
\end{document}